\documentclass{emulateapj}

\countdef\decade=200
\advance\decade by \year
\countdef\hours=201
\advance\hours by \time
\divide\hours by 60
\countdef\mins=202
\advance\mins by \hours
\multiply\mins by 60
\multiply\hours by 100
\countdef\miltime=203
\advance\miltime by \hours
\advance\miltime by \time
\advance\miltime by -\mins

\shorttitle{Line Emission From Massive Disks}
\shortauthors{Krumholz, Klein, \& McKee}

\newcommand{\ltsim}{\protect\raisebox{-0.5ex}{$\:\stackrel{\textstyle <}
        {\sim}\:$}}
\newcommand{\gtsim}{\protect\raisebox{-0.5ex}{$\:\stackrel{\textstyle >}
        {\sim}\:$}}

\newcommand{\msun}{M_{\odot}}
\newcommand{\lsun}{L_{\odot}}

\begin{document}

\title{Molecular Line Emission from Massive Protostellar Disks:\\Predictions for ALMA and the EVLA}

\slugcomment{Accepted for publication in the Astrophysical Journal, April 23, 2007}

\author{Mark R. Krumholz\altaffilmark{1}}
\affil{Department of Astrophysical Sciences, Princeton University,
Princeton, NJ 08544}
\email{krumholz@astro.princeton.edu}

\author{Richard I. Klein}
\affil{Astronomy Department, University of California, Berkeley,
Berkeley, CA 94720, and Lawrence Livermore National Laboratory,
Livermore, CA 94550}
\email{klein@astron.berkeley.edu}

\author{Christopher F. McKee}
\affil{Departments of Physics and Astronomy, University of California,
Berkeley, Berkeley, CA 94720}
\email{cmckee@astron.berkeley.edu}

\altaffiltext{1}{Hubble Fellow}

\begin{abstract}
We compute the molecular line emission of massive protostellar disks by solving the equation of radiative transfer through the cores and disks produced by the recent radiation-hydrodynamic simulations of Krumholz, Klein, \& McKee. We find that in several representative lines the disks show brightness temperatures of hundreds of Kelvin over velocity channels $\sim 10$ km s$^{-1}$ wide, extending over regions hundreds of AU in size. We process the computed intensities to model the performance of next-generation radio and submillimeter telescopes. Our calculations show that observations using facilities such as the EVLA and ALMA should be able to detect massive protostellar disks and measure their rotation curves, at least in the nearest massive star-forming regions. They should also detect significant sub-structure and non-axisymmetry in the disks, and in some cases may be able to detect star-disk velocity offsets of a few km s$^{-1}$, both of which are the result of strong gravitational instability in massive disks. We use our simulations to explore the strengths and weaknesses of different observational techniques, and we also discuss how observations of massive protostellar disks may be used to distinguish between alternative models of massive star formation.
\end{abstract}


\section{Introduction}

Determining the nature of the disks or other structures that feed gas into massive protostars is one of the major observational challenges in radio and submillimeter astronomy. Observations to date have revealed large-scale toroids in sub-Keplerian rotation, and in a few cases evidence for Keplerian rotation (see recent reviews by \citealt{cesaroni06, cesaroni07a}, \citealt{beuther06b}, and \citealt{beuther07a}). However, even the most sensitive observations thus far have only been able to probe length scales of thousands to tens of thousands of AU from the embedded massive star. True accretion disks are likely to begin at smaller distances from their host stars, so determining the nature of massive stellar disks and their role in the star formation process will require yet higher resolution.

Observationally, this presents a significant challenge. Even the closest regions of massive star formation are $\sim 0.5$ kpc away, so resolving structures hundreds of AU in size requires tenth-of-an-arcsecond angular resolution. A more typical distance for massive protostellar cores is $1-2$ kpc, requiring even higher resolution. Since the strong ionizing flux from a massive star rapidly alters or destroys its natal environment, it is only possible to observe the structures from which a massive protostar forms while it is still embedded and accreting. Such embedded stars are generally obscured by hundreds of magnitudes of dust extinction at visual wavelengths, and can therefore only be observed at radio or submillimeter wavelengths. Furthermore, fully understanding the behavior of these structures requires kinematic information, which can only be obtained unambiguously from line emission. Probing molecular lines requires significantly higher sensitivity than continuum observations.

Unfortunately, no existing telescopes are capable of detecting molecular line emission at brightness temperatures hundreds of Kelvin or less with resolutions $\sim 0."1$. However, this combination of sensitivity and resolution is within reach of next-generation observatories such as the EVLA and ALMA, and studying the environments in which massive protostars form is one of the major science goals of these facilities. Fulfilling this goal will require significant theoretical input, both to suggest what observing strategies are likely to be successful, and to make predictions which, when compared with observations, can be used to help decide between alternative models.

Recently, \citet[hereafter KKM07]{krumholz07a} simulated the early stages of collapse and fragmentation of massive protostellar cores. These simulations are ideal for observational comparisons for several reasons. First, they begin from realistic, turbulent initial conditions, and thus should have levels of structure and complexity comparable to real massive star formation regions. In comparison, many other simulations of massive star formation begin from far simpler initial conditions with no turbulence \citep[e.g.][]{yorke02,banerjee07a}. While these simulations are an excellent tool for isolating and exploring some of the physical problems of massive star formation, their simplicity makes them less suited for detailed observational comparison than more complex models.

Second, KKM07 include radiative transfer as well as hydrodynamics and gravity in their simulations, including radiative heating by the forming protostar(s). As a result, the simulations make realistic predictions for the distribution of temperatures, as well as densities and velocities, inside a star-forming core. In contrast, models that neglect radiation and treat the gas using only a modified equation of state \citep[e.g.][]{bonnell04, dobbs05}, or that include radiation only via cooling curves \citep[e.g.][]{banerjee06,banerjee07a}, do not produce realistic temperature distributions because they neglect heating by accreting protostars \citep[KKM07]{krumholz06b}.

In this paper we use the simulations of KKM07 to predict the molecular line emission of embedded, accreting massive protostars. We focus on molecular lines because they are the only available means of  following gas kinematics, and therefore of establishing whether observed structures truly are disks. In \S~\ref{method}, we describe our computational method for making this prediction. This section is rather technical, and is likely to be of greatest interest to computational specialists. In \S~\ref{results}, we present simulated emission data for a selection of molecular lines. We also process the data through simple models to mimic the performance of next-generation telescopes. In \S~\ref{discussion} we consider the implications of our results for future observations, including how these observations may be used to distinguish between competing models of massive star formation. We also discuss the limitations of our model and how these are likely to impact our predictions. Finally, in \S~\ref{conclusion} we summarize our conclusions. 

\section{Computational Methodology}
\label{method}

\subsection{Calculation of the Emergent Intensity}
\label{calcmethod}

\subsubsection{Ray Tracing Through an Adaptive Grid}

\begin{figure}
\plotone{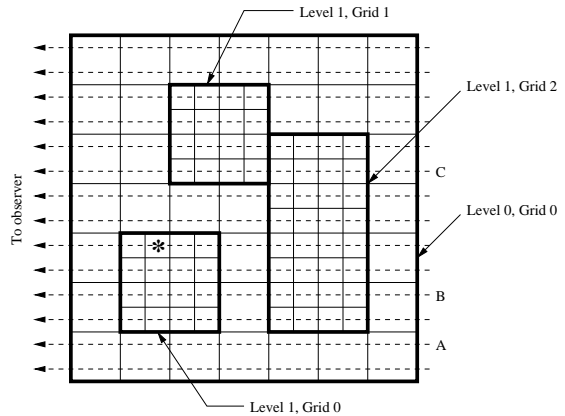}
\caption{\label{amrraytrace}
Illustration of rays (\textit{dashed lines}) passing through an AMR grid. Cell edges are indicated by thin solid lines, and grid boundaries are indicated by thick solid lines. Cells are numbered by (\textit{column}, \textit{row}) from the lower left corner of a grid, starting with cell 0, so the point marked by the asterisk is in cell $(1,3)$ of level 1, grid 0, and also cell $(1,2)$ of level 0, grid 0. The cell list for ray $A$ is level 0, grid 0, cells $(0,0),\ldots,(6,0)$. The list for ray $B$ is level 0, grid 0, cell $(0,1)$; level 1, grid 0, cells $(0,1),\ldots,(3,1)$; level 0, grid 0, cell $(3,1)$; level 1, grid 2, cells $(0,1),\ldots,(3,1)$; level 1, grid 0, cell $(6,1)$. The cell list for ray $C$ is level 0, grid 0, cells $(0,4),(1,4)$; level 1, grid 1, cells $(0,0),\ldots,(3,0)$; level 1, grid 2, cells $(0,6),\ldots,(3,6)$; and level 0, grid 0, cell $(6,4)$.
}
\end{figure}

We wish to compute the radiative intensity emerging from a region within which information on the density, velocity, and temperature fields is recorded on an adaptive grid consisting of a series of levels $l=0,1,\ldots L$ on which the cell size is $\Delta x_l$. The coarsest, lowest resolution level is $l=0$, and the highest resolution level is $l=L$. To compute the intensity from this data, we first construct a grid of rays, representing different lines of sight from an observer at infinity, passing through the computational domain. The rays are evenly spaced in a rectangular grid on the plane at infinity, with adjacent rays separated by a distance $\Delta x_L$, where $\Delta x_L$ is the length of the finest cell in the adaptive grid. Along each ray, we construct a list of the cells through which it passes, ordered so that cell 0 is closest to the observer. At every point in space, we record in the list only the finest resolution cell covering that point. Figure \ref{amrraytrace} illustrates this procedure.

From the cell list for a given ray, we construct the list of densities $\rho_0,\ldots \rho_n$, line-of-sight velocities $v_0,\ldots, v_n$, and temperatures $T_0,\ldots,T_n$ through which the ray passes, with subscripts increasing away from the observer, and velocities defined so that positive corresponds to motion away from the observer. We also have the list of the lengths of the ray segments $ds_0,\ldots ds_n$ intersecting each cell. We treat the density, velocity, and temperature fields as step functions, so that at distance $s$ along the ray such that $s_i<s<s_{i+1}$, the density, velocity, and temperature are $\rho_i$, $v_i$, and $T_i$, where $s_i\equiv \sum_{j=0}^i ds_j$, and $s=0$ corresponds to the point at which the ray enters the computational domain.

\subsubsection{Emissivities and Absorption Coefficients}

Now that we know the density $\rho$, line-of-sight velocity $v$, and temperature $T$ along each ray, we must use these to compute the emissivity and absorption coefficient along the ray. We consider two emission and absorption mechanisms. The first is dust thermal emission and absorption. Following \citet{chakrabarti05}, we adopt a modified version of the \citet{draine03a,draine03b,draine03c} opacity law (which is an updated version of the \citealt{weingartner01} law). We set the dust specific absorption coefficient $\kappa_{\nu}^{\rm dust}$ at frequency $\nu$ equal to that predicted by Draine's model for $R_V=5.5$ in gas with a temperature above $110$ K. In colder gas we set the absorption coefficient to double that of the Draine model to account for the extra opacity provided by ice mantles on dust grains. {\it Note that $\kappa_{\nu}^{\rm dust}$ is the opacity per unit gas mass, not the opacity per unit dust mass.} Thus, the total absorption coefficient is simply $\alpha^{\rm dust}_{\nu}=\kappa_{\nu}^{\rm dust} \rho$, and since the dust emits thermally, the emissivity is $j_{\nu}^{\rm dust}=\alpha_{\nu}^{\rm dust} B_{\nu}(T)$, where $B_{\nu}(T)$ is the Planck function. At the radio and submillimeter wavelengths with which we are concerned, scattering is negligible and we need not distinguish between absorption and extinction.

The second source of emission and absorption is molecular lines. Consider a line emitted by a species $X$ transition from an upper state $u$ to a lower state $l$. The line is at a frequency $\nu_{ul}$ in the frame comoving with an emitting atom. In this paper we limit ourselves to species and levels that are in local thermodynamic equilibrium (LTE). Since the density in the massive cores we wish to examine is everywhere larger than $\sim 10^6$ cm$^{-3}$, and in the disk it is typically $\sim 10^{10}$ cm$^{-3}$, this is not a significant restriction. We discuss the validity of the LTE approximation for individual species in \S~\ref{results}. For a species in LTE, the number density of molecules of species $X$ in state $l$ is
\begin{equation}
n_{l} = f_X g_l \frac{\rho}{\mu_H} \left[\frac{\exp\left(-\frac{E_{l}}{k_B T}\right)}{Q(T)}\right],
\end{equation}
where $f_X$ is the abundance of species $X$ relative to hydrogen, $g_l$ is the statistical weight of state $l$, $\mu_H$ is the mean gas mass per H nucleus ($=2.34\times 10^{-24}$ g for a gas of hydrogen and helium in the standard cosmic abundance), $E_l$ is the energy of state $l$ relative to the ground state, and $Q(T)$ is the partition function for species $X$. The corresponding absorption coefficient at frequency $\nu$ is
\begin{equation}
\alpha_{\nu}^{\rm line} = \frac{h\nu}{4\pi} n_l B_{lu} \left[1-\exp\left(-\frac{E_{ul}}{k_B T}\right)\right] \phi(\nu),
\end{equation}
where $B_{lu}$ is the Einstein coefficient for absorption from state $l$ to state $u$, $E_{ul}$ is the energy difference between states $u$ and $l$, and $\phi(\nu)$ is a function describing the line shape. 
In LTE, the source function is simply the Planck function, so the emissivity
\begin{equation}
j_{\nu}^{\rm line} = \alpha_{\nu}^{\rm line} B_{\nu}(T).
\end{equation}

To determine the line shape function, we assume that the gas has a Maxwellian velocity distribution, and that both the bulk velocity and the velocity dispersion are piecewise-constant along the line of sight. We ignore the intrinsic width of the line relative to the thermal and bulk-motion broadening. If the bulk velocity of the gas is $v$, then a photon of frequency $\nu$ in the observer's frame has frequency $\nu'=(1+\beta) \nu$ in the frame comoving with the gas, where $\beta=v/c$. The velocity dispersion of species $X$ along the line of sight is $\sigma_v=\sqrt{k_B T/m_X}$, where $m_X$ is the mass of particles of species $X$. The corresponding dispersion in frequencies is $\sigma_{\nu} = (\sigma_v/c)\nu_{ul}$. Combining the bulk velocity and the velocity dispersion of the gas, the line shape function is
\begin{equation}
\phi(\nu) = \frac{1}{\sqrt{2\pi \sigma_{\nu}^2}} \exp\left\{
-\frac{
\left[(1+\beta)\nu-\nu_{ul}\right]^2
}{
2\sigma_{\nu}^2
}
\right\}.
\end{equation}

In practical calculations it is generally more convenient to work in velocity offsets relative to the mean velocity of the gas than in frequencies. If we define an observation at velocity $v_{\rm obs}$ to mean an observation at a frequency $\nu=(1-v_{\rm obs}/c)\nu_{ul}$, then we may alternately write the line shape function as
\begin{equation}
\phi(v_{\rm obs}) = \frac{1}{\sqrt{2\pi \sigma_{\nu}^2}} \exp\left[
-\frac{(v-v_{\rm obs})^2}{2\sigma_v^2}
\right].
\end{equation}
For all the calculations presented in this paper, we perform our computations over velocities ranging from $-20$ to $20$ km s$^{-1}$ relative the velocity of the primary star (see \S~\ref{simdata}), with a velocity resolution of $0.1$ km s$^{-1}$. Since $0.1$ km s$^{-1}$ smaller than the thermal velocity dispersion of the molecules we discuss below even in the coolest parts of our computational domain, this resolution is sufficient to sample the emission well.

Our assumption that the velocity is piecewise constant is potentially problematic, since in reality there are velocity gradients along the line of sight that correspond to velocity differences from one side of a cell to another that are not necessarily small compared to the thermal sound speed of the gas. Failure to account for these gradients in our calculations causes the line profile we compute emerging from a given pixel to be somewhat jagged. However, we are interested in aggregate quantities that average over many cells and not in detailed line shapes. Since the jaggedness in our spectra does not significantly change integrated quantities such as the total intensity or intensity-weighted mean velocity at a given pixel, we do not attempt to correct for it by including a treatment of line of sight gradients.

\subsubsection{Solution of the Transfer Equation}

Now that we have determined the emissivity and the absorption coefficient, the final step is to solve the transfer equation,
\begin{equation}
\label{transfereq}
\frac{dI_{\nu}}{ds} = -j_{\nu} + \alpha_{\nu} I_{\nu},
\end{equation}
where $I_{\nu}$ is the radiative intensity at frequency $\nu$ traveling in the $-s$ direction, and $j_{\nu}=j_{\nu}^{\rm dust}+j_{\nu}^{\rm line}$ and $\alpha_{\nu} = \alpha_{\nu}^{\rm dust}+\alpha_{\nu}^{\rm line}$ are the total emissivity and absorption coefficient from dust and lines. Note that taking $I_{\nu}$ to be positive in the $-s$ direction inverts the usual sign convention, and this is the reason our transfer equation also has the reverse of the usual sign convention. 

Fortunately, since the emissivity and the absorption coefficient are functions simply of density, velocity, and temperature, and these are piecewise-constant on our grid, $j_{\nu}$ and $\alpha_{\nu}$ are likewise piecewise-constant. This makes solution of the transfer equation trivial. For ray segment $i$, within which the emissivity is $j_{\nu,i}$ and the absorption coefficient is $\alpha_{\nu,i}$, integrating the transfer equation (\ref{transfereq}) from $s_i$ (the boundary between segments $i$ and $i+1$) to $s_{i-1}$ (the boundary between segments $i-1$ and $i$) gives
\begin{equation}
\label{transfersegment}
I_{\nu}(s_{i-1}) = \exp(-\tau_{\nu,i}) I_{\nu}(s_i)  + [1-\exp(-\tau_{\nu,i})] B_{\nu,i},
\end{equation}
where $\tau_{\nu,i}=\alpha_{\nu,i} ds_i$ is the optical depth through segment $i$, $B_{\nu,i}$ is the Planck function evaluated at the temperature in segment $i$, and $I_{\nu}(s_i)$ is the intensity entering the segment. Combining a series of integrations of the form (\ref{transfersegment}), and taking the intensity entering the computational domain from the direction opposite the observer to be zero, the intensity emerging to reach the observer from $s=0$ is simply
\begin{equation}
I_{\nu}(0) = \sum_{i=0}^n \exp\left(-\sum_{k=0}^{i-1} \tau_{\nu,i}\right) \left[1 - \exp(-\tau_{\nu,i})\right] B_{\nu,i}.
\end{equation}

\subsection{Input Molecular Data}

We obtain the molecular data required for the computation described in \S~\ref{calcmethod} from the Jet Propulsion Laboratory Molecular Spectroscopy Catalog\footnote{http://spec.jpl.nasa.gov/} \citep{pickett98}. For each species, the catalog provides the value of the partition function at temperatures of 9.375, 18.75, 37.5, 75, 150, 225, and 300 K. We compute our partition function at other temperatures by fitting a piecewise-powerlaw function to the catalog values. At temperatures above 300 K, we extrapolate using the same powerlaw index as between 225 and 300 K. We take the abundances of each species we use from values available in the literature, as described below in \S~\ref{results}.

\subsection{Input Simulation Data}
\label{simdata}

\begin{figure*}
\plotone{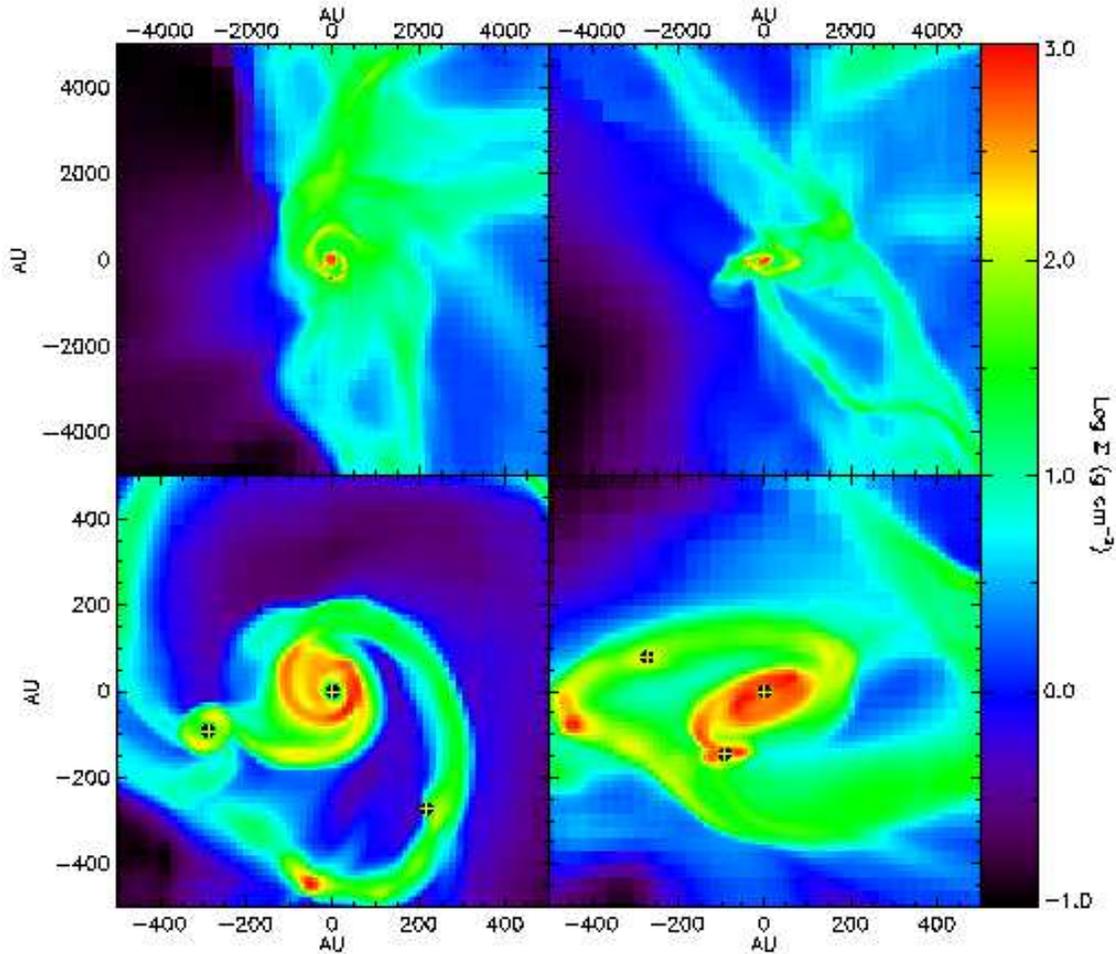}
\caption{\label{truecol}
Column density in regions $10000$ AU and $1000$ AU in size (\textit{upper and lower rows}) in two projections (\textit{left and right columns}). Stars are shown in the $1000$ AU images by plus signs. The $8.3$ $\msun$ star is at the origin of both plots. In the image on the left, the $1.8$ $\msun$ star is to the one to the left of the origin; in the right image, the $0.05$ $\msun$ star is the one on the far left. The overdense blob attached to the disk spiral arm, shown near the bottom of the bottom left image and the left of the bottom right image, is in the process of collapsing to form an additional fragment, but has not collapsed yet. Note that the drop in column density in the immediate vicinity of the $8.3$ and $1.8$ $\msun$ stars is an artifact of the finite resolution of the simulation, not a true physical effect.\\
}
\end{figure*}

We apply our radiative transfer calculations to the density, velocity, and gas temperature fields produced by the simulations of KKM07. We only briefly review the simulation methodology and initial conditions here, and refer readers to KKM07 and the companion methodology paper \citet{krumholz07b} for a full description. Although the radiative transfer method we describe in this paper is applicable to any grid-based simulation, all the predicted emission line data we present here comes from applying this method to run 100A of KKM07, so we focus on it in our description. This run begins from a spherical core with a mass of 100 $\msun$ and an initial radius of $0.1$ pc. The core is initially centrally condensed, with density profile $\rho^{-1.5}$ at large radii, and a maximum initial density of $10^{-14}$ g cm$^{-3}$ at radii $<38.4$ AU. The core has an initial turbulent velocity field with a velocity dispersion of 1.7 km s$^{-1}$ chosen to put the initial core into approximate hydrostatic balance. The turbulent power spectrum is $P(k)\propto k^{-2}$. The initial temperature in the core is 20 K, and the radiation boundary condition in the simulation is that there is a uniform radiative flux equal to that of a 20 K blackbody radiation field entering the computational domain. Radiation generated within the domain may escape freely at the domain edge.

KKM07 evolve the core using the Orion code, which solves the equations of radiation-hydrodynamics on an adaptive mesh in the gray, flux-limited diffusion approximation. The simulation domain is a cube 0.6 pc on side, and the peak resolution is $7.5$ AU, a dynamic range of $16,384$ cells per linear dimension. KKM07 report the results of the first 20 kyr of evolution, but we have continued the simulations using the same code, and we present simulated maps based on the state of the core after 27 kyr of evolution. At this time, the core has formed a primary protostar of $8.3$ $\msun$, surrounded by a disk $3-5$ $\msun$ in mass, and $500-1000$ AU in radius. The exact mass and size of the disk are somewhat indeterminate, since in the simulation the disk has no sharply-defined edge. The central star has completely burned its accumulated deuterium reserve but has not yet contracted onto the main sequence. Through a combination of Kelvin-Helmholtz contraction and accretion power it is producing a time-averaged luminosity of approximately $10^{3.5}-10^4$ $\lsun$. The luminosity and mass of the central object make the simulated core an analog of systems such as IRAS 20126+4104 \citep{zhang99, cesaroni05}. In addition to the primary star, there are two other stars in the simulation domain that are $1.8$ and $0.05$ $\msun$ in mass, neither of which is producing much luminosity in comparison to the primary star. We show the column density in the simulation at this time in Figure \ref{truecol}. Note that, viewed in either orientation, the column density through the disk reaches almost $1000$ g cm$^{-2}$.

Before using the simulation data for our radiative transfer calculations, we modify it in two ways. First, in the simulation KMM07a impose a constant pressure boundary condition by surrounding the core with an ambient medium within which the temperature is 100 times higher and the density is 100 times lower than the gas at the edge of the core. To avoid contaminating the predicted intensities with contributions from this medium, we set the density to zero in any cell within which the density is less than $5$ times the initial ambient medium density. Since this is a factor of 20 lower than the initial core edge density, this threshold should not eliminate a significant amount of core material.

The second modification is to the temperature field. In any given time step there are usually a handful of cells where the iterative radiation solver in Orion has not converged, and the temperature is unphysical. While these cells have no significant effect on the dynamics, since they disappear after a single time step and never contain a significant amount of mass, unphysically warm cells can affect the predicted intensities. A single cell that is twice as hot as its neighbors can outshine them even if it contains very little mass, particularly in velocity channels close to that cell's velocity. We therefore remove these unphyiscally warm cells as follows. The signature of the non-convergence is a rapid oscillation in temperature over a few cells. We detect this signature by flagging as suspect any cell where, along the line-of-sight for which we are computing the transfer, there is both a local maximum and a local minimum in the temperature within 6 computational cells. We replace the temperature in such flagged cells with a linear interpolation of the temperatures of the neighboring non-flagged cells. Visual inspection of the modified temperature fields that we produce using this procedure shows that they agree well with what one would identify and correct by eye.

As a final point, although in this paper we only present radiative transfer calculations for run 100A of KKM07, we have processed other simulations from KKM07 using the same techniques. The details of the predicted line emission obviously differ from run to run, since the underlying density and temperature distributions differ in detail. However, we find that there is no significant qualitative difference between the line emission predicted for run 100A and that predicted for other runs at times when the primary star's mass is comparable to $8.3$ $\msun$, the mass of the primary in run 100A at the time we consider.

\section{Predicted Molecular Line Emission}
\label{results}

\subsection{Choice of Lines}

In principle we can use the method described in \S~\ref{method} to compute the predicted emission from any molecule in LTE. At the densities of $\sim 10^9$ cm$^{-3}$ or more typical of the structures we are interested in tracing, this includes most species. However, selection of appropriate species is not trivial.
It is only possible to obtain kinematic information if the emission at a given frequency is dominated by molecular emission rather than dust emission, or the two are at least comparable. Since the intensity emitted by the disk along a given line of sight cannot exceed the value of a blackbody, and if the dust is optically thick it will radiate as a blackbody, it is only possible for molecular emission to dominate if the dust is optically thin. This is a significant challenge: at 100 GHz, the opacity predicted by our dust model is $\kappa_{\nu}^{\rm dust} = 5.6\times 10^{-4}$ cm$^2$ g$^{-1}$, and the opacity in cold regions where grains have ice mantles will be $1.1\times 10^{-3}$ cm$^2$ g$^{-1}$. (Again, note that this is the opacity per unit gas mass, not the opacity per unit dust mass.) Figure \ref{truecol} indicates that the column density in the inner disk, within $\sim 100$ AU of the primary star, falls in the range $100-1000$ g cm$^{-2}$. This means that the inner disk has optical depth of at least a few tenths, usually more, in dust, and thus is it is not possible to obtain velocity information for it at 100 GHz or higher frequencies.

This calculation suggests that the best tool for tracing very small scales is radio observations at frequencies well below 100 GHz, where the dust contribution is smallest. Unfortunately, the brightness temperature sensitivity to line emission for current and future radio telescopes such as the VLA and EVLA is much poorer than the sensitivity of current and future submillimeter telescopes such as PdBI, the SMA, and ALMA. Thus, the hot inner disk is likely to be the \textit{only} part of the structure around our protostar detectable at radio frequencies using the high resolutions required to probe the structure of the disk, and this will likely require very long integration times. Observations at frequencies of 100 GHz or more are far more efficient for tracing somewhat lower column density structures on larger scales.

\begin{figure*}
\plotone{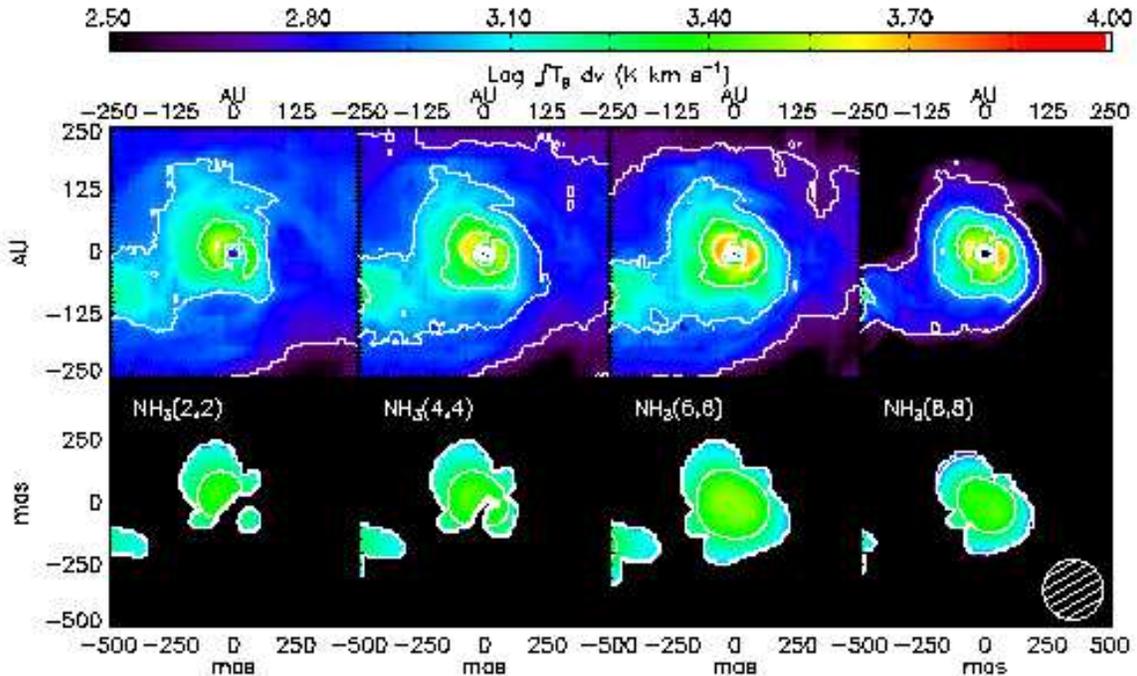}
\caption{\label{nh3faceimage}
For the disk seen face-on, velocity-integrated brightness temperature emitted in the indicated NH$_3$ inversion transitions (\textit{upper row}), and integrated brightness temperature detected in a simulated observation of these lines using the EVLA (\textit{lower row}). In the simulated observation, we show as black any pixel for which there is no detection at the $3\sigma(=144\mbox{ K})$ level in any velocity channel. The beam size for the simulated observation is shown by the hatched region in the lower right-hand corner. Contours start at 500 K km s$^{-1}$, and increase by factors of 2 per contour.\\
}
\end{figure*}

Based on these considerations, we present simulated observations for the lines of two molecules. At radio frequencies we use the inversion transitions of ammonia (NH$_3$), and in the submillimeter regime we use some of the strongest rotational transitions of methyl cyanide (CH$_3$CN). Both molecules have previously been used to trace structures around massive stars on larger scales and at lower resolutions with the current generation of telescopes (see the recent reviews by \citealt{cesaroni06, cesaroni07a}, \citealt{beuther06b}, and \citealt{beuther07a}), and thus are known to be present and excited in massive protostellar envelopes. We simulate observations for distances of 0.5 kpc and 2 kpc. The former is roughly the distance to the Orion molecular cloud, the nearest region of massive star formation, while the latter is a more typical for observations of massive star-forming regions. We assume that all telescope beams are Gaussian, and do not include the removal of large-scale power by interferometers in our models.

\subsection{Ammonia}

\begin{figure*}
\plotone{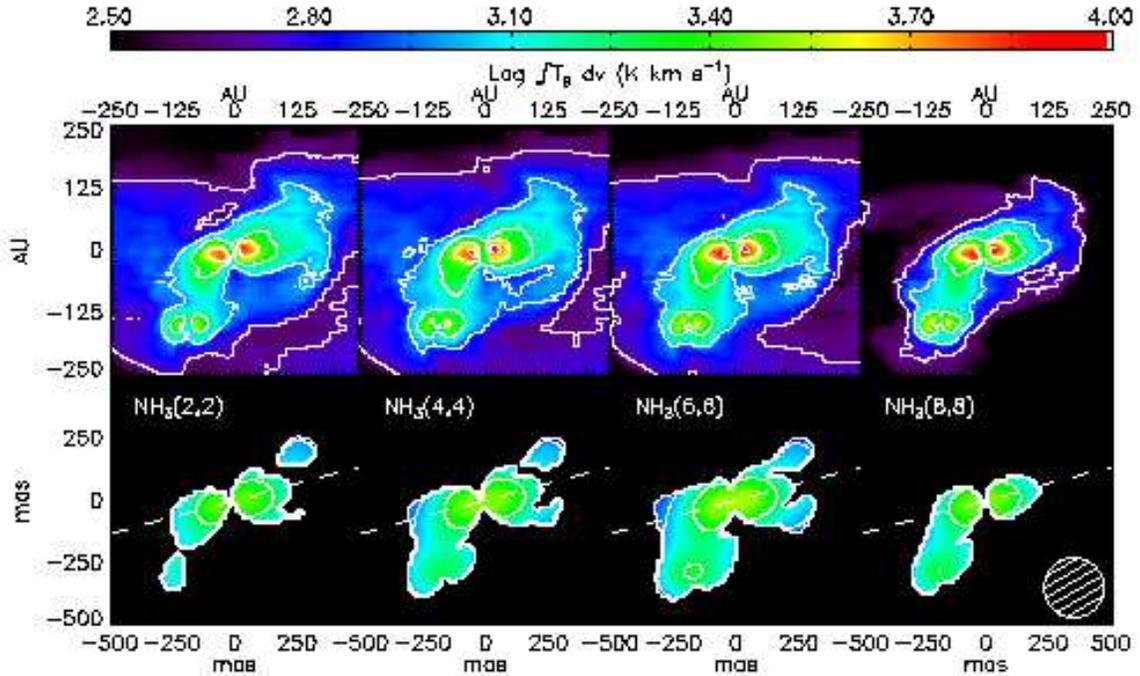}
\caption{\label{nh3edgeimage}
For the disk seen edge-on, velocity-integrated brightness temperature. See the caption of Figure \ref{nh3faceimage} for details of what is shown in each panel. The dashed lines in the simulated observations indicate the orientation of the disk, inclined $15^{\circ}$ with respect to the $x$ axis in the observation.
}
\end{figure*}

\begin{figure*}
\plotone{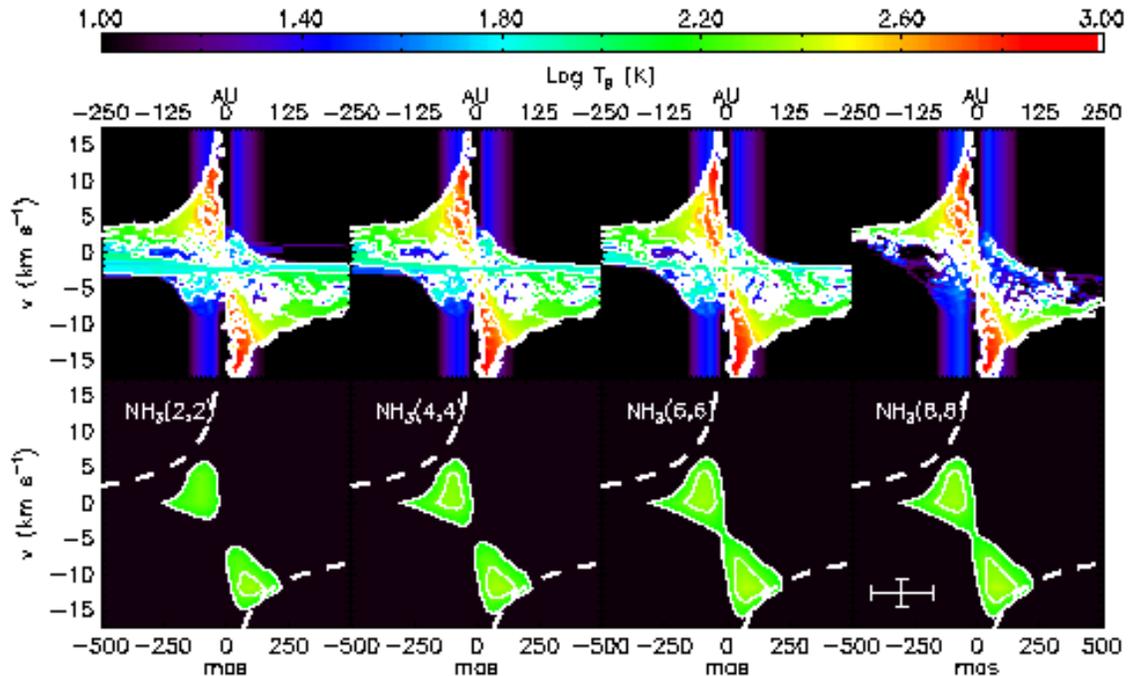}
\caption{\label{nh3pv}
Brightness temperature in the indicated NH$_3$ inversion transitions as a function of position and velocity, with position measured relative to the star along the dashed line indicated in Figure \ref{nh3edgeimage}. We show both the true brightness temperature (\textit{upper row}) and a simulated EVLA observation (\textit{lower row}). The contours start at 50 K and increase by factors of 2. In the simulated observation, we report 0 for any pixel in which the signal is smaller than the $3\sigma(=144\mbox{ K})$ noise level. The angular and velocity resolution of the observations are indicated by the error bars in the lower right-hand image. The thick dashed lines indicate the velocity profile of a Keplerian orbit around a star of $8.3$ $\msun$, offset by a constant velocity of $-3$ km s$^{-1}$ relative to the zero velocity in the image.
}
\end{figure*}

We first compute the radiation intensity emerging from a massive protostellar core in the $(J,K)=(2,2)$, $(4,4)$, $(6,6)$, and $(8,8)$ inversion transitions of NH$_3$, at frequencies of 23.7226, 24.1394, 25.0560, and 26.5190 GHz. The upper levels for these transitions have temperatures of 65.0, 201.1, 408.6, and 687.4 K above the ground state, so they probe a wide range of temperatures, including those that are expected to occur only in the inner disk very close to the star. At temperatures $\ge 20$ K, the critical densities for the upper states of these transitions are all $\ltsim 10^3$ cm$^{-3}$. Since the density everywhere in our core is $\gtsim 10^6$ cm$^{-3}$, our assumption of LTE for these levels is well-justified. (To compute the critical densities, we use the molecular data available in the Leiden Atomic and Molecular Database\footnote{http://www.strw.leidenuniv.nl/$\sim$moldata/}, \citealt{schoier05}, since the JPL database does not include the collision strengths required to compute critical densities.)

The abundance of NH$_3$ in dense molecular cores is fairly uncertain. In low mass, cold cores observations show that ammonia abundance rises with density \citep{tafalla04a, tafalla04b}, reaching values of $10^{-6}$ at the high densities found in our core. Observations of massive star-forming regions generally find significantly lower abundances of $10^{-8}-10^{-7}$ \citep{harju93, tieftrunk98}. However, these abundances are based on single-dish observations of distant regions that average over size scales of $\gtsim 0.5$ pc, and  thus presumably include a great deal of low density material with correspondingly low ammonia abundance. Indeed, \citet{harju93} find that their derived ammonia abundances decrease systematically with distance, a result they attribute to a decrease in the NH$_3$ filling factor as their fixed-angular size beam averages over larger and larger spatial scales. It therefore seems likely that the NH$_3$ abundance in our disk is significantly higher than the values derived from large-scale averaging, and we adopt the larger value of $10^{-6}$ determined for low mass cores as our fiducial abundance. Fortunately, our results are fairly insensitive to this choice, because the NH$_3$ line is optically thick through the bulk of the disk unless the abundance is more than an order of magnitude lower than our fiducial value.

Finally, note that our calculations are only for the main part of the NH$_3$ line, not the hyperfine satellites. These satellites are intrinsically weaker by a factor of $\sim 14$. While this would decrease their optical depths and thus make them more faithful tracers of kinematics, as we show below even the optically thick NH$_3$ main line emission is difficult to detect even in long integrations. These weaker lines would be harder still to observe.

Figures \ref{nh3faceimage} and \ref{nh3edgeimage} show the velocity-integrated intensity in the four NH$_3$ lines for our core at a distance of 0.5 kpc viewed nearly face-on and nearly edge-on, in the orientations shown in the left and right columns of Figure \ref{truecol}. In both Figures, the upper row shows the velocity-integrated brightness temperature as a function of position relative to the primary star. As one might expect, the low-temperature $(2,2)$ line comes from a large region extending hundreds of AU away from the central star, while the $(8,8)$ line, which requires temperatures of nearly 700 K to be excited, comes almost entirely from a region within 100 AU of the primary and secondary stars. Also note that the disk around the $1.8$ $\msun$ secondary appears quite distinctly, and that there is also a hot spot visible around the $0.05$ $\msun$ star.

The lower rows of Figures \ref{nh3faceimage} and \ref{nh3edgeimage} show simulated observations of the disks using the EVLA phase I, which at the frequencies of the NH$_3$ inversion lines is projected to have a FWHM resolution of $0."12$ in the A configuration. We adopt  a velocity resolution of 4 km s$^{-1}$ for the simulated observation. The estimated 1 $\sigma$ thermal noise of the EVLA in A configuration for a 12 hour integration in K-band with this bandwidth is 48 K \citep{perley01}. We take the $1\sigma$ noise in the integrated intensity to be the noise in a channel times the channel width, $192$ K km s$^{-1}$. Note that, although the EVLA is capable of far finer velocity resolution, simulated observations with significantly smaller velocity channels produce images with very low signal to noise ratios; since 4 km s$^{-1}$ resolution is sufficient to map the rotation of the inner disk, we adopt it. With these observational parameters, the system is reasonably well resolved, and the presence of a flattened disk is obvious in the edge-on observation. The secondary is also clearly resolved as a separate peak in the in the edge-on images, although the material around it is not clearly shown as a disk due to beam smearing. The tertiary star is outside the box shown, and the gas around it is below the sensitivity limit of the observation in both orientations.

Figure \ref{nh3pv} shows the true brightness temperature and a simulated EVLA observation as a function of velocity and position along the disk viewed edge-on, indicated by the dashed line in Figure \ref{nh3edgeimage}. As in the total intensity map, the disk is reasonably well-detected and marginally resolved in all four transitions. The velocity profile of the disk is well-fit by a Keplerian orbit around an $8.3$ $\msun$ star, the true mass of the central object. Interestingly, however, in order to obtain a good fit to the true emission, the mean velocity of the orbit must be shifted by $-3$ km s$^{-1}$ relative to the velocity of the star, which is zero on the axes shown. However, the offset persists even in the true brightness temperature map. Its physical origin is in the fact that the star is really moving with respect to the disk, due to the large mass of the disk relative to the star. This is part of the gravitational instability mechanism for angular momentum transport which is operating in the disk, as discussed \citet{laughlin94} and KKM07. In the simulated observations a somewhat larger offset would be required to obtain a good fit, but the difference is below the velocity resolution limit of the simulated observation. 

Whether this offset between the central velocity of the disk and the stellar velocity is detectable observationally depends on whether it is possible measure the velocity of the star. Since the star is deeply embedded, one cannot measure its velocity directly. However, if such an embedded star produces a relatively small ultraviolet flux, it may generate a hypercompact HII region which will be detectable in radio emission through the intervening gas. During the hypercompact phase when ionized gas is confined very close to the stellar surface, the ionized gas velocity should presumably be close to the stellar velocity. The 8.3 $\msun$ star we model here probably produces too little ionizing flux to generate a detectable HII region, however.

We also compute simulated observations for a distance of 2 kpc. However, at this distance the entire region of NH$_3$ emission detectable in a 12 hour integration falls within a single beam pointing at the EVLA's highest resolution. For this reason, we do not discuss these simulated observations further.

\subsection{Methyl Cyanide}

\begin{figure*}
\epsscale{0.85}
\plotone{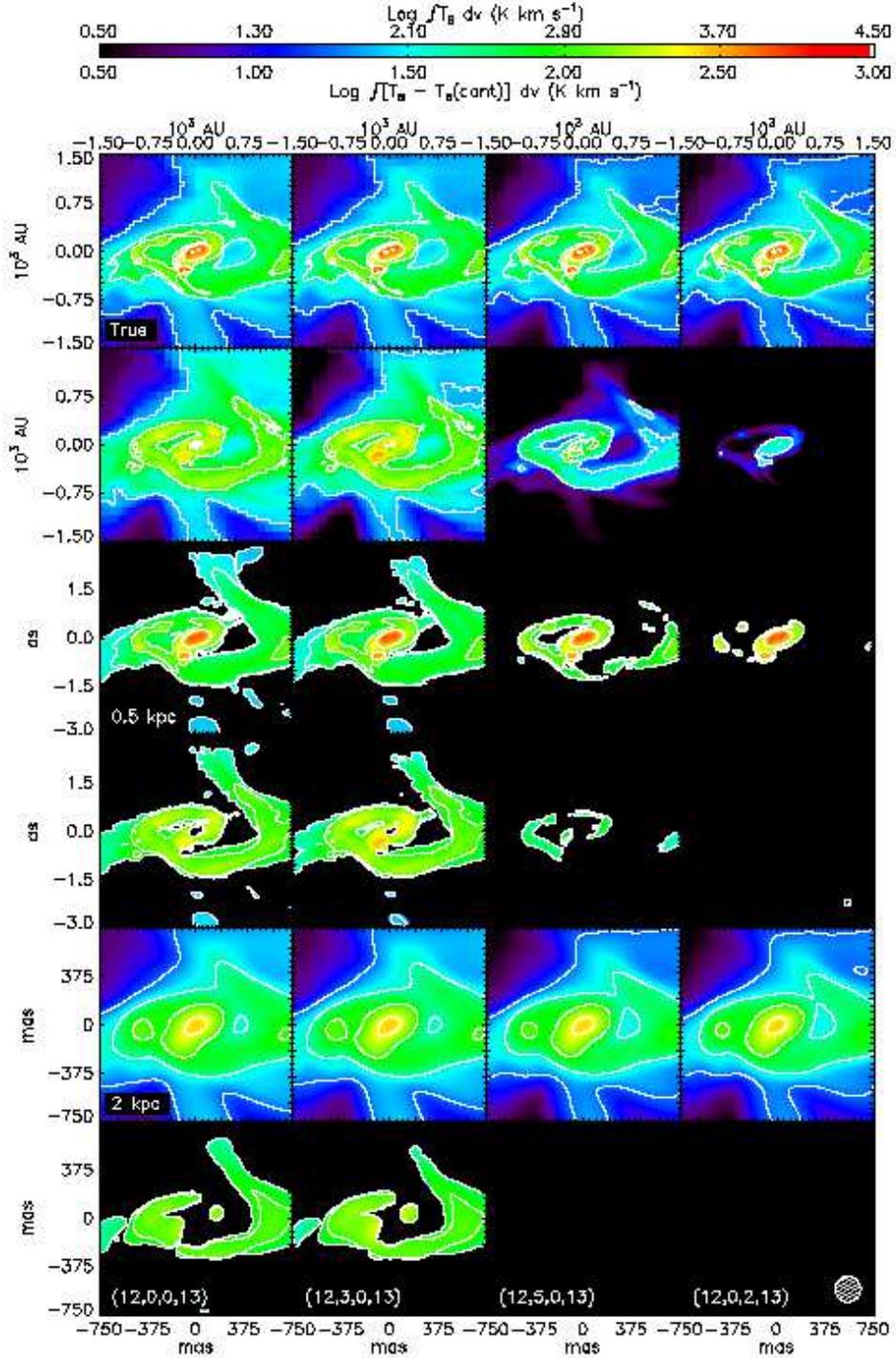}
\epsscale{1.0}
\caption{\label{ch3cnedgeimage}
Velocity-integrated brightness temperature emitted in the indicated CH$_3$CN rotational transitions over a 40 km s$^{-1}$ velocity range centered about the center of each line. We show the true total emission (\textit{upper row}), the true continuum-subtracted emission (\textit{second row from the top}), the total emission as observed with ALMA (\textit{third row from the top}) at 0.5 kpc, the continuum-subtracted emission as observed with ALMA (\textit{fourth row from the top}), and simulated ALMA observations of the total and continuum-subtracted emission at 2 kpc (\textit{fifth and sixth rows, respectively}). The upper level of the transition shown in each image is indicated in each column. The size of the ALMA beam at 0.5 kpc and 2 kpc is indicated in the lower right corners of the fourth and sixth rows. In the true continuum-subtracted emission, a black pixel indicates a position where the line is seen in absorption rather than emission. In the simulated observations, black pixels correspond to positions where there is no detection at the $3\sigma(=33\mbox{ K})$ level in any velocity channel. For the total emission images (\textit{odd rows}), the contours start at 50 K km s$^{-1}$; in the continuum-subtracted images (\textit{even rows}), they start at 25 K km s$^{-1}$. In all cases they increase by factors of 4 per contour. Note the difference in scales from Figures \ref{nh3faceimage} and \ref{nh3edgeimage}, and also note that the color scales are different for the total and continuum-subtracted images.
}
\end{figure*}

We next compute the emission from our disk in four rotational transitions of CH$_3$CN: $(12,0,0,13)\rightarrow (11,0,0,12)$, $(12,3,0,13)\rightarrow (11,3,0,12)$, $(12,5,0,13)\rightarrow (11,5,0,12)$, and $(12,0,2,13)\rightarrow (11,0,2,12)$. Here the quantum numbers are given in the JPL catalog format $(N,K,v,J)$ for symmetric rotors with vibration, where $N$ is the total angular momentum of the molecule excluding electron spin, $K$ is the projection of this onto the molecule's symmetry axis, $v$ is the vibrational quantum number, and $J$ is the total angular momentum including electron spin. These transitions occur at frequencies of 220.7472, 220.7091, 220.6411, and 221.3942 GHz, and their upper states lie 69.0, 133.3, 247.6, and 594.6 K above the ground state. We choose these transitions because they are among the strongest for the molecule, with Einstein $A$'s of almost $10^{-3}$ s$^{-1}$.

Unfortunately collision strengths for the CH$_3$CN levels we examine are not available in the astrophysical literature, so we cannot compute critical densities. However, collision strengths for lower CH$_3$CN levels \citep{pei95} give critical densities $\ltsim 10^7$ cm$^{-3}$ even for the strongest radiative transitions, as do collision strengths for molecules of similar complexity such as methanol (CH$_3$OH). (Data for methanol are from the Leiden Atomic and Molecular Database, \citealt{schoier05}.) Since we are concerned with emission from the disk and structures around it, where typical densities are generally $10^8$ cm$^{-3}$ or more, it seems reasonable to assume that methyl cyanide molecules are in LTE.

The abundance of CH$_3$CN is, unfortunately, even less well understood than that of NH$_3$. The formation of methyl cyanide is driven by complex organic chemistry that occurs at the comparatively high temperatures of hot cores like the one in our simulation. There is strong evidence that the abundance of nitrogenous organic molecules is time-dependent \citep{beuther06b}. Rather than attempt to add a chemistry model to our radiative transfer calculation, we adopt a fiducial CH$_3$CN abundance of $10^{-8}$ independent of density. This is roughly consistent with the values measured by \citet{hatchell98} along lines of sight toward 14 ultracompact HII regions. We discuss how variations in the abundance will affect our results below.

Figure \ref{ch3cnedgeimage} shows the velocity-integrated brightness temperature in the four CH$_3$CN transitions seen edge-on. As with Figures \ref{nh3faceimage} and \ref{nh3edgeimage}, the upper row shows the true brightness temperature. Since dust thermal emission dominates essentially everywhere when integrated over the 40 km s$^{-1}$ interval for which we compute the output, this is effectively a dust continuum image integrated over a 40 km s$^{-1}$ channel. To isolate the molecular line emission (and in some cases absorption) we subtract from each bin the emitted intensity in the $-20$ km s$^{-1}$ bin, which contains a negligible contribution from line emission, in order to produce a continuum-subtracted image. We show this in the second row of Figure \ref{ch3cnedgeimage}. The inner disks around each star are mostly seen in absorption or weak emission, while the larger-scale extended structure shows strong line emission.

\begin{figure*}
\plotone{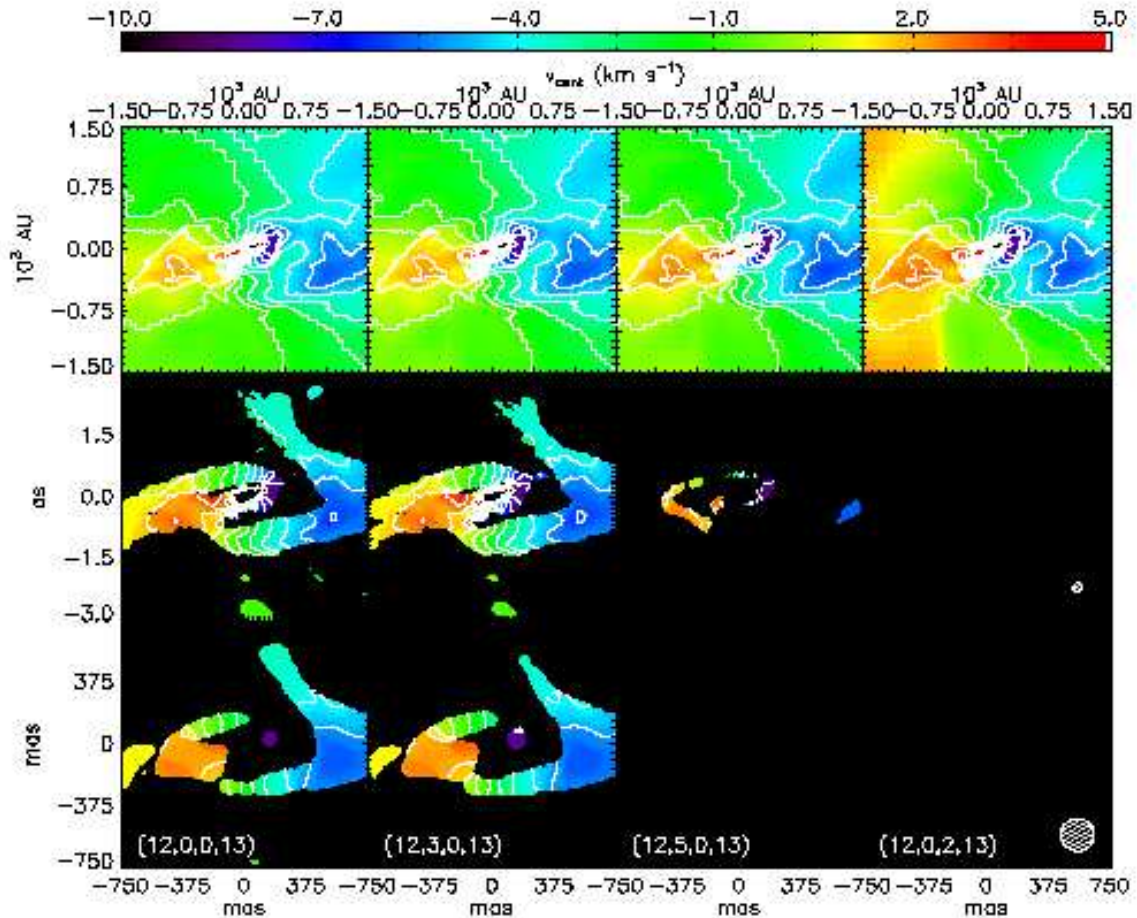}
\caption{\label{ch3cnedgepv}
Centroid velocity as a function of position in the indicated CH$_3$CN rotational transitions, after subtracting off dust continuum emission. We show the true centroid velocity (\textit{top row}) and the result of a simulated observation with ALMA assuming a distance of 0.5 kpc (\textit{middle row}) and 2.0 kpc (\textit{bottom row}). The contours are separated by intervals of 2 km s$^{-1}$. The thick contours correspond to negative velocities, and the thin contours correspond to positive velocities.\\
}
\end{figure*}

Note that our continuum-subtraction procedure is only a rough approximation to the continuum subtraction procedure used in real interferometric observations, which involves fitting the observed visibilities in the uv-domain with polynomials. However, the qualitative conclusions we draw about the effects of continuum-subtraction do not depend in detail on this point.

The bottom four rows show simulated observations of the total and continuum-subtracted images made with ALMA, at distances both 0.5 kpc and 2 kpc. For the simulated ALMA observations, we adopt a velocity resolution of 0.5 km s$^{-1}$, an angular resolution of $0."1$, and assume 1000 seconds per pointing. With these values, the $1\sigma$ noise level is roughly 11 K per channel for each line. (We compute the noise level using the ALMA sensitivity calculator\footnote{http://www.eso.org/projects/alma/science/bin/\\sensitivity.html} for 50 antennas.)

As expected, in all four transitions the inner disk, within $\sim 100$ AU of the primary star, is detected in absorption of the dust continuum by molecules in cooler foreground gas. Absorption information is difficult to use to determine disk kinematics, because there is no easy way to separate absorption arising from a cool surface layer of the disk, which would trace its motion, from absorption arising from foreground gas elsewhere in the core that is not part of the disk. As a result, we will not attempt to use the line detected in absorption to trace kinematics of the inner disk. On the other hand, the outer disk at larger radii is well-detected and resolved in emission in two of the four lines, and marginally detected in the third. The highest temperature line provides no significant signal because the region of the core hot enough to emit it is largely confined to the region where the dust is optically thick and the line is detected primarily in absorption.

This calculation also illustrates the importance of resolution and distance. With infinite resolution, as indicated in the second row of Figure \ref{ch3cnedgeimage}, the entire disk is visible in line emission. With lower resolution, however, it is not possible to subtract off the dust emission faithfully and recover the line emission on size scales comparable to the beam size or smaller. The optically thick continuum emission of the inner disk and the gas immediately around the star raises the emitted intensity to the blackbody intensity, and since this is the maximum possible emitted intensity in LTE, the line cannot be seen above it. Consequently, less and less of the disk becomes observable as the distance, and thus the physical scale probed by a beam of a given angular size, increases. In order to probe the inner disk as close to the star as possible, one therefore wants the highest possible angular resolution.

In the transitions where the disk is detected, the strong spiral structure of the disk is readily apparent at either distance. One can confirm that the structure is a spiral arm in a disk rotating about the central object by examining its velocity structure. Figure \ref{ch3cnedgepv} shows the true value and a simulated ALMA observation of the centroid velocity of the continuum-subtracted line emission. We define the centroid velocity as the continuum subtracted-brightness temperature-weighted velocity average at each point. In computing the average for the true brightness temperature, we only include pixels for which the continuum subtracted-brightness temperature is positive, for the true value map, or is greater than the $3\sigma$ noise level of the beam, for the simulated observation.

Although it is not directly possible to compare the rotation profile shown by Figure \ref{ch3cnedgepv} with the Keplerian speed, since the disk is clearly non-circular and the distance from the central object to the point in the arm from which emission arises is not clearly determined, the simulated observation clearly does reveal rotation consistent with what one would expect for a spiral arm in a disk. At $500-1500$ AU from the primary 8.3 $\msun$ star, the Keplerian speed is $2.2-3.8$ km s$^{-1}$, consistent with the observed velocities for a distance of 0.5 kpc. At an assumed distance of 2 kpc, the velocities are considerably lower, due to blending of material at different velocities. Also note that both the true brightness temperature map and the simulated observation reveal significantly more negative velocity than positive velocity gas; the entire structure is shifted away from zero velocity. This is the same offset seen in the NH$_3$ observation on smaller scales. However, it is important to remember that the centroid velocities are not necessarily perfectly reliable tracers of the true velocity, particularly toward the denser parts of the spiral arm, because the lines becomes optically thick in places.

\begin{figure*}
\plotone{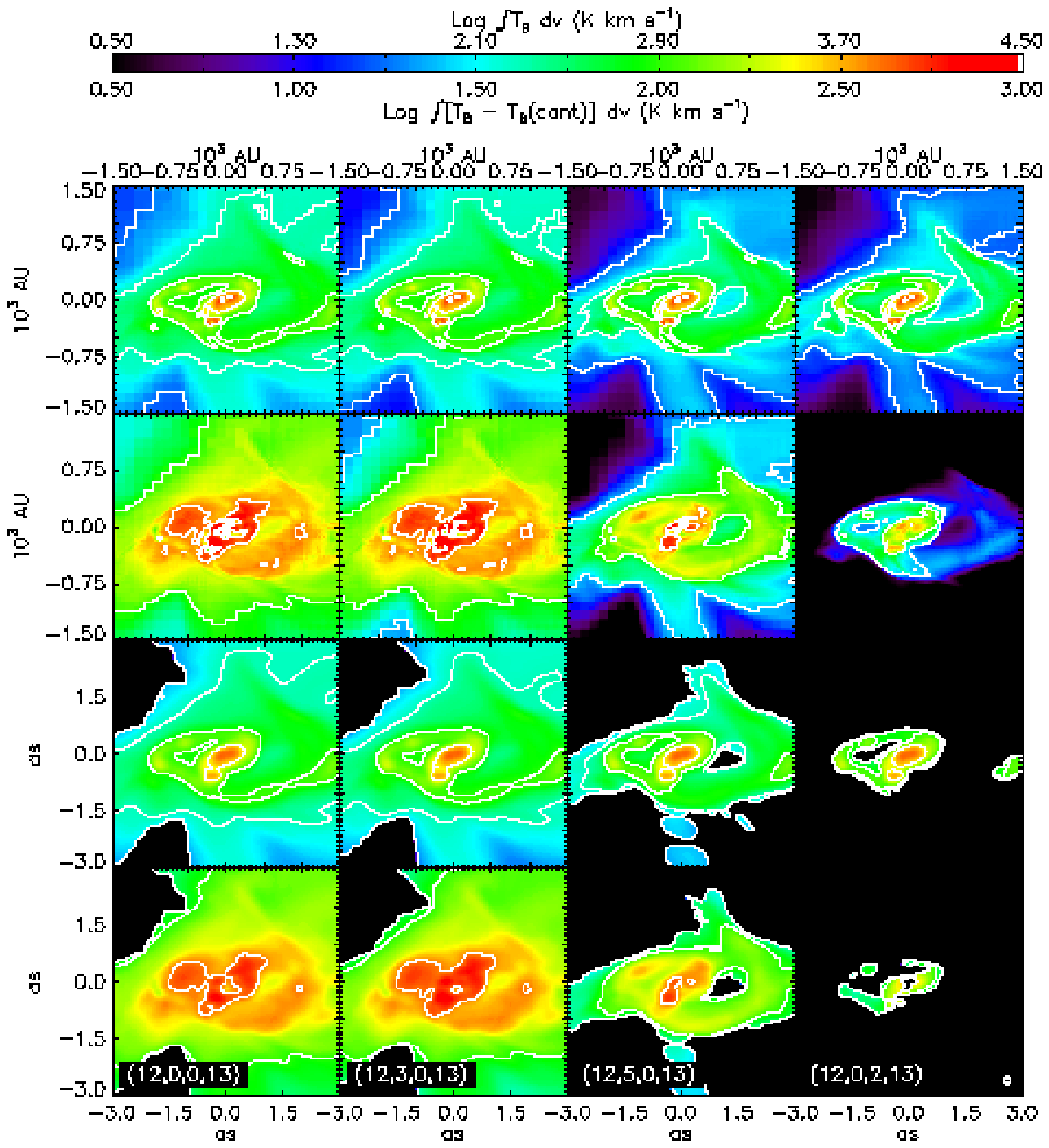}
\caption{\label{ch3cnvaredgeimage}
Same as the top four rows of Figure \ref{ch3cnedgeimage}, but with a CH$3$CN abundance of $10^{-7}$.
}
\end{figure*}

\begin{figure*}
\plotone{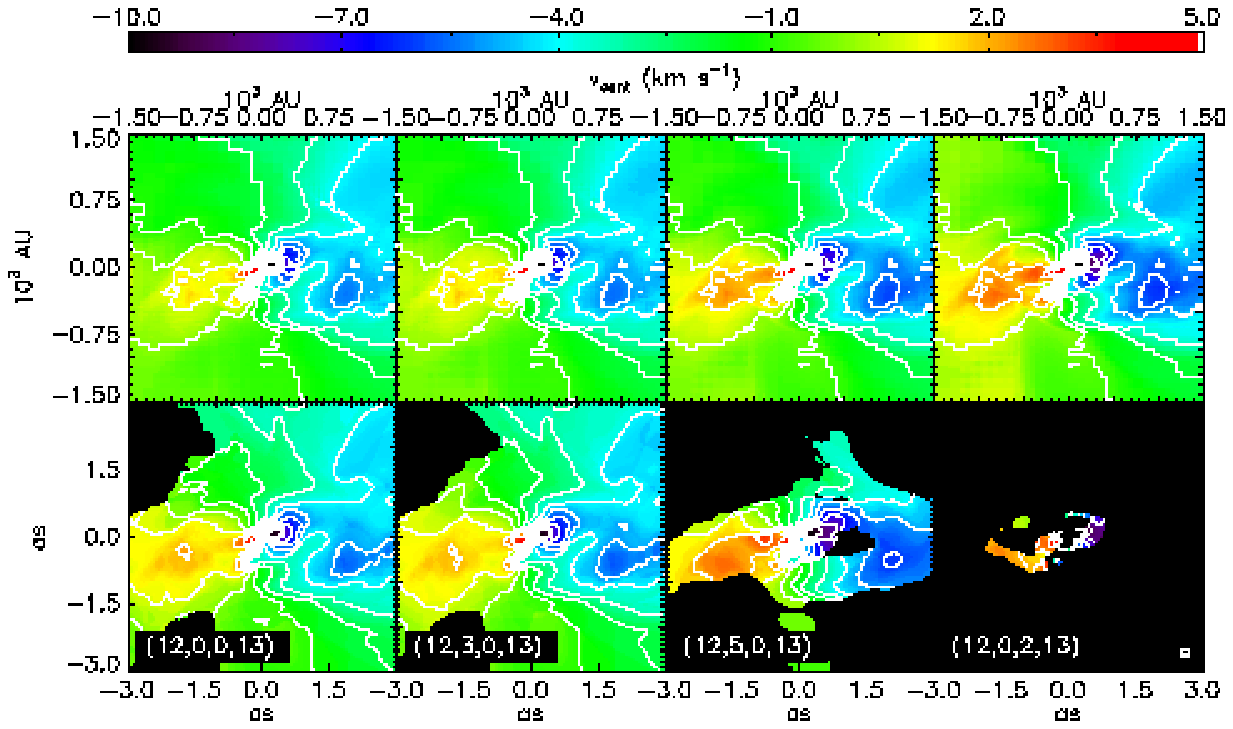}
\caption{\label{ch3cnvaredgepv}
Same as the top four rows of Figure \ref{ch3cnedgepv}, but with a CH$3$CN abundance of $10^{-7}$.
}
\end{figure*}

Finally, we return to the question of how much our results depend on the assumed abundance of methyl cyanide. To help explore this question, we calculate emission from the same four lines, adopting abundances a factor of 10 higher and lower than our ``best guess" value of $10^{-8}$. For simplicity, we only compare calculations at 0.5 kpc, not 2 kpc. Not surprisingly, we find that our results depend fairly strongly on the abundance. With an abundance of $10^{-9}$, we find almost no detectable continuum-subtracted line emission using the same sensitivity and resolution as for our calculations with an abundance of $10^{-8}$. In contrast, figures \ref{ch3cnvaredgeimage} and \ref{ch3cnvaredgepv} show the velocity-integrated intensity and centroid velocity as a function of position for an abundance of $10^{-7}$. In this case, we find line emission within ALMA's sensitivity arising from a significantly larger region, and find detectable emission after subtracting the continuum from all four lines, although the detection for the highest temperature line is over a fairly small area. However, the emission in this case traces the kinematics less faithfully than for an abundance of $10^{-8}$, particularly for the lower levels, because the line is quite optically thick in many regions. For example, the peak positive velocity in the $(12,0,0,13)$ and $(12,3,0,13)$ lines is noticeably lower than the peak positive velocity in those lines for an abundance of $10^{-8}$, or than the peak positive velocity in the $(12,5,0,13)$ line for an abundance of $10^{-7}$.

\section{Discussion}
\label{discussion}

\subsection{Implications for Future Observations}

Our analysis shows that the $\sim 1000$ AU-sized disks around young, accreting massive protostars should be detectable and at least marginally resolvable with the EVLA and ALMA at distances of 0.5 kpc, the distance to the nearest region of massive star formation. Perhaps most excitingly, these detections should unambiguously reveal rotation at velocities comparable to the Keplerian speed, and should also reveal the presence of large spiral arms, or any other large-scale non-axisymmetry. They will therefore enable us for the first time to probe the structure of the accretion disks that feed gas onto massive protostars. However, these observations will be costly. On the $\sim 0."1$ size scales resolvable by these telescopes, the brightness temperature is hundreds of Kelvin, and obtaining strong detections of these signals will require tens of hours of integration.

Our calculations also reveal some of the strengths and weaknesses of radio versus submillimeter observations. The column density in the inner parts of disks around massive protostars can reach of order 1000 g cm$^{-2}$, viewed either edge-on or face-on. This column density is high enough to make dust optically thick at submillimeter wavelengths, making it impossible to detect line emission and trace disk kinematics for the inner parts of disks using submillimeter observations. Only radio observations are able to trace the kinematics in the inner parts of disk. However, because existing and planned radio telescopes have relatively little sensitivity to lines, they are only likely to be able to detect the very inner parts of disks with reasonable integration times. In contrast, the greater sensitivity of submillimeter telescopes makes them excellent tools for tracing the outer, lower column density parts of disks. However, one does require high angular resolution in order to minimize the amount of the outer disk concealed by the bright, optically thick dust emission from the inner disk.

Our results on the optical thickness of massive disks are in interesting contrast with observations of disks around low mass protostars. Models of these disks based on multi-wavelength observations generally indicate that dust reaches optical depths of unity only at frequencies $\gtsim 350$ GHz and radii of $\ltsim 50$ AU \citep[e.g.][]{lay97}. The difference in our results arises simply from the much greater column densities that massive disks can reach. Even class 0 low mass protostellar disks typically have $\Sigma \ltsim 100$ g cm$^{-2}$ at radii larger than $100$ AU, an order of magnitude lower than the column densities reached in our disks at similar radii.

\subsection{Distinguishing Models of Massive Star Formation}

One compelling reason to search for disks around massive protostars, and to explore their properties, is to distinguish between competing models of massive star formation. The two dominant models are core accretion \citep[KKM07]{mckee02, mckee03}, in which massive stars form by the collapse of massive gas cores directly to massive stars, and collision / competitive accretion \citep{bonnell98,bonnell03,bonnell05}, in which all stars are born with masses of $\sim 0.5$ $\msun$, and massive stars reach their final mass by a combination of runaway gas accretion and star-star collisions in the centers of stellar clusters. The evidence for each of these models have recently been reviewed by a number of authors, including \citet{krumholz06f}, \citet{bonnell07a}, and \citet{beuther07a}.

High resolution observations of the environments of massive protostars offer a strong method of distinguishing between the models, and more generally a method of determining the nature of the massive star formation process. In particular, the coalescence/competitive accretion model predicts that essentially all stars larger than 3 $\msun$ should suffer close encounters with other stars, truncating their disks to sizes smaller than $\sim 30$ AU \citep{bonnell03}. The core accretion model, as exemplified by the KKM07 simulation, predicts that massive stars should have disks a few hundred to a few thousand AU in size, with masses that are typically tens of percent of the mass of the primary star. These disks may also have other protostars embedded within them or around them \citep[see also][]{kratter06}, which may or may not be detectable as ``hot spots" in the disks, depending on the separation and orientation. Thus, the detection of structures such as those presented in this paper would be very strong evidence for core accretion, while non-detections with ALMA and the EVLA would be strong evidence for competitive accretion.

\subsection{Limitations of Our Calculations}

Our predictions are subject to several limitations. Most significantly is our poor knowledge of the abundances of tracer molecules, and how these change in time and space. As discussed above, there is considerable evidence that the abundances of both NH$_3$ and CH$_3$CN vary depending on the ambient density and temperature, and on the time history of density and temperature experienced by a particular fluid element. \citet{beuther06b} describes recent observational work toward developing a chemical evolutionary sequence for hot molecular cores, but these efforts are still preliminary.  Consequently, our abundance estimates could easily be off significantly. This would have strong effects on observations, as is clear from comparison of Figures \ref{ch3cnedgeimage} with \ref{ch3cnvaredgeimage}, and \ref{ch3cnedgepv} with \ref{ch3cnvaredgepv}. Due to this limitation it is probably best to regard our predictions as examples of what lines in the radio and submillimeter regimes may look like. For a given source, the particular tracer molecules we have used may be much more or less abundant, in which case one should be able to obtain data comparable to what we present using alternative molecules with transitions at similar frequencies and upper level temperatures. In the absence of a reliable, predictive chemodynamical model, the only option may be to try many molecules until a useful one is detected.

Another limitation is in the KKM07 simulation that we use to produce our predicted line emission. The details of the simulation and its approximations and limitations are discussed in detail in KKM07, but two in particular are worth mentioning here because of their potentially large impact on our predicted line emission. First, the KKM07 simulations do not contain protostellar outflows, which are often the most prominent features in some tracers. These are obviously missing from our predictions, and thus we cannot comment on potential confusions in determining whether a given observational feature is tracing disks or outflows.

Second, the KKM07 simulations begin with isolated, spherical cores, and thus are likely to differ from real cores on scales comparable to the initial core radius, $0.1$ pc. On these scales, the geometry of the core at formation and interactions between the core and its environment are likely to have significant effects on both morphology and kinematics. For this reason, we do not expect our cores to reproduce observations on scales of that are significant fractions of $0.1$ pc. However, while gas outside the core may affect the morphology and kinematics of the core gas, it is important to point out that this gas will
{not in itself present a significant source of emission or opacity. Massive cores are generally embedded in molecular clumps with column densities $\sim 1$ g cm$^{-2}$ \citep[e.g.][]{plume97}, so we expect this to be the typical column density of any intervening material between the core and the observer. At 100 GHz, this corresponds to a dust optical depth $\sim 10^{-3}$, completely negligible in either absorption or emission. The line opacity will also be negligible in the lines we examine in this paper, because the lower states for these lines have excitation temperatures $\gtsim 60$ K. Since the gas in molecular clumps is typically at much lower temperatures of $\sim 20$ K \citep{plume97}, there will be a negligible number of molecules excited into the lower states of our chosen transitions, and thus almost no molecules capable of absorbing photons coming from the core.

\section{Conclusions}
\label{conclusion}

Using detailed radiation-hydrodynamic simulations of massive star formation as a starting point, we predict the molecular line emission of the disks around massive, embedded protostars. These disks are distinct from the massive, rotating toroids discussed by \citep{cesaroni06} and others, in that they are Keplerian rather than sub-Keplerian, they are geometrically thin, and they are much smaller in both spatial extent and mass. We find that the disks emit line radiation at brightness temperatures ranging from hundreds of kelvins within a few hundred AU of the central star, down to many tens of kelvins at distances of $\sim 1000$ AU. This emission will occur over velocity channels $10-20$ km s$^{-1}$ wide. These brightness temperatures and bandwidths should make line emission from massive protostellar disks detectable and at least marginally resolvable with next generation telescopes such as ALMA and the EVLA.

Disks around young, massive protostars have extremely high column densities in their centers, and the resulting high dust columns require radio observations to probe structures and kinematics within $100-200$ AU of the central star. At submillimeter wavelengths it is possible to image disks in dust thermal emission, but not to obtain kinematic information from lines. Line observations in the radio over distances of a few hundred AU should reveal clear evidence for disks rotating at speeds comparable to the Keplerian speed around the central object. The zero velocity for the Keplerian rotation curve may be offset relative to the velocity of the central star by a few km s$^{-1}$ due to relative motion of the disk and star induced by strong gravitational instability in the star-disk system, although it will only be possible to detect this offset if one can observationally determine the stellar velocity.

Submillimeter telescopes, with their greater sensitivities, are more suited to mapping the outer, extended parts of disks between a few hundred and $1000-2000$ AU from the primary star. Observations on these scales should reveal if massive protostellar disks have significant spiral structure, as the simulations of KKM07 predict they should. They should also show clear evidence for rotation. However, even in the outer disk, obtaining kinematic information requires observation of particularly strong lines and/or abundant molecules in order for the emission to be detectable in reasonable integration times. It will also require high spatial resolution so that as little line emission from the inner disk as possible is masked by bright continuum emission from close to the protostar.

Observations of disks around massive protostars can play a decisive role in deciding between competing models for the formation of massive stars. The core accretion model predicts that massive protostars should be surrounded by disks $\sim 500-1500$ AU in size, with masses that are appreciable fractions of the mass of the central protostar. In contrast, the competitive competitive accretion / collision model predicts an absence of such massive, extended disks. Given the potential power of high-resolution radio and submillimeter observations for probing the nature of the star formation process, we encourage all simulators and modelers of star formation to make predictions that can be directly compared to observations using the next generation of telescopes.

\acknowledgements
We thank H. Beuther and the anonymous referee for detailed comments on the manuscript that improved the quality of the paper.  We also thank D. Backer, M. Beltran, B. Draine, A. Goodman, C. Heiles, B. Parise, and E. Rosolowsky for helpful suggestions and illuminating discussions about radio astronomy and astrochemistry. Support for this work was
provided by NASA through Hubble Fellowship grant \#HSF-HF-01186
awarded by the Space Telescope Science Institute, which is operated by
the Association of Universities for Research in Astronomy, Inc., for
NASA, under contract NAS 5-26555 (MRK); NASA ATP grants NAG 5-12042
and NNG06GH96G (RIK and CFM); the US Department of Energy at the
Lawrence Livermore
National Laboratory under contract W-7405-Eng-48 (RIK); and the NSF
through grants AST-0098365 and AST-0606831 (CFM). This research was
also supported by grants of high performance computing resources from
the Arctic Region Supercomputing Center; the NSF San Diego
Supercomputer Center through NPACI program grant UCB267; the National
Energy Research Scientific Computing Center, which is supported by the
Office of Science of the U.S. Department of Energy under Contract
No. DE-AC03-76SF00098, through ERCAP grant 80325; and the US
Department of Energy at the Lawrence Livermore National Laboratory
under contract W-7405-Eng-48.


\end{document}